\documentclass[conference]{IEEEtran}
\IEEEoverridecommandlockouts

\usepackage{cite}
\usepackage{amsmath,amssymb,amsfonts}
\usepackage[acronym]{glossaries}
\usepackage{graphicx}
\usepackage{textcomp}
\usepackage{xcolor}
\usepackage[caption=false,font=footnotesize]{subfig}
\usepackage{algorithm}
\usepackage{algpseudocode}
\usepackage{listings}
\usepackage{breqn}
\usepackage{authblk}
\usepackage{hyperref}
\def\BibTeX{{\rm B\kern-.05em{\sc i\kern-.025em b}\kern-.08em
    T\kern-.1667em\lower.7ex\hbox{E}\kern-.125emX}}
\newtheorem{lemma}{Lemma}
\newacronym{dd}{DD}{delay-Doppler}
\newacronym{otfs}{OTFS}{orthogonal time frequency space}
\newacronym{idft}{IDFT}{inverse discrete Fourier transform}
\newacronym{dft}{DFT}{discrete Fourier transform}
\newacronym{dt}{DT}{delay-time}
\newacronym{cp}{CP}{cyclic prefix}

\newacronym{2d}{2D}{two dimensional}
\newacronym{dtft}{DTFT}{discrete time Fourier transform}
\newacronym{tf}{TF}{time-frequency}
\newacronym{esprit}{ESPRIT}{estimation of signal parameters via rotational invariance techniques}
\newacronym{los}{LoS}{line of sight}
\newacronym{nlos}{NLoS}{non line of sight}
\newacronym{isfft}{ISFFT}{inverse symplectic finite Fourier transform}
\newacronym{sfft}{SFFT}{symplectic finite Fourier transform}
\newacronym{fft}{FFT}{fast Fourier transform}
\newacronym{ifft}{IFFT}{inverse fast Fourier transform}
\newacronym{1g}{1G}{first generation}
\newacronym{4g}{4G}{fourth generation}
\newacronym{5g}{5G}{fifth generation}
\newacronym{6g}{6G}{sixth generation}
\newacronym{ckm}{CKM}{channel knowledge map}
\newacronym{aod}{AoD}{angle of departure}
\newacronym{mse}{MSE}{mean square error}
\newacronym{ep}{EP}{embedded pilot}
\newacronym{bs}{BS}{base station}
\newacronym{fl}{FL}{federated learning}
\newacronym{cds}{CDS}{cyclic difference set}
\newacronym{csi}{CSI}{channel state information}
\newacronym{uav}{UAV}{unmanned aerial vehicle}
\newacronym{mmimo}{mMIMO}{massive multiple-input multiple-output}
\newacronym{mmwave}{mmWave}{millimeter-wave}
\newacronym{ofdm}{OFDM}{orthogonal frequency-division multiplexing}
\newacronym{imi}{IMI}{identical multi-set intersection}
\newacronym{isi}{ISI}{inter-symbol interference}
\newacronym{cs}{CS}{compressed sensing}
\newacronym{siso}{SISO}{single-input-single-output}
\newacronym{iai}{IAI}{inter-antenna interference}
\newacronym{ici}{ICI}{inter-carrier interference}
\newacronym{rf}{RF}{radio frequency}
\newacronym{fr2}{FR2}{frequency range 2}
\newacronym{fr1}{FR1}{frequency range 1}
\newacronym{3gpp}{3GPP}{3rd Generation Partnership Projec}
\newacronym{ompfr}{OMPFR}{orthogonal matching pursuit with fractional
	refinement}
\newacronym{ss}{SS}{spatially spread}
\newacronym{doa}{DoA}{direction of arrival}
\newacronym{dda}{DDA}{delay, Doppler, and angular}
\newacronym{3d}{3D}{three dimensional}
\newacronym{pgm}{PGM}{proximal gradient descent}
\newacronym{ls}{LS}{least square}
\newacronym{lms}{LMS}{least mean square}
\newacronym{kf}{KF}{Kalman filter}
\newacronym{ekf}{EKF}{extended Kalman filter}
\newacronym{bem}{BEM}{basis expansion model}
\newacronym{ser}{SER}{symbols error rate}
\newacronym{ukf}{UKF}{unscented Kalman filter}
\newacronym{crlb}{CRLB}{Cramer-Rao lower bound}
\newacronym{vkf}{VKF}{vector Kalman filter}
\newacronym{scm}{SCM}{spatial
	channel model}
\newacronym{cir}{CIR}{channel impulse response}
\newacronym{satc}{SATC}{spatial average of temporal correlation}
\newacronym{ar}{AR}{autoregressive}
\newacronym{mlp}{MLP}{multi-layer perceptron}
\newacronym{lmmse}{LMMSE}{linear minimum mean square error}
\newacronym{gdss}{GDSS}{gabor division spread spectrum}
\newacronym{rmse}{RMSE}{root mean square error}
\newacronym{zblp}{ZBLP}{zero-bin low-PAPR}
\newacronym{papr}{PAPR}{peak-to-average-power ratio}
\newacronym{zpotfs}{ZP-OTFS}{zero-pad OTFS}
\newacronym{rv}{RV}{range-velocity}
\newacronym{ftst}{FTST}{fast-time-slow-time}

\newacronym{idi}{IDI}{inter-Doppler interference}
\newacronym{vrcf}{VRCF}{variable resolution coarse-to-fine}
\newacronym{isac}{ISAC}{integrated sensing and communication}
\newacronym{mlmfs}{ML-MFS}{maximum likelihood based on multi-frame superposition}
\newacronym{woamfs}{WOA-MFS}{whale optimization multi-frame superposition}
\newacronym{ml}{ML}{maximum likelihood}
\newacronym{am}{A\&M}{Aboutanios and Mulgrew}
\newacronym{haqse}{HAQSE}{hybrid A\&M and q-shift estimator}
\newacronym{cfh}{CFH}{coarse-to-fine HAQSE}
\newacronym{music}{MUSIC}{Multiple Signal Classification}
\newacronym{snr}{SNR}{signal-to-noise ratio}
\newacronym{oddm}{ODDM}{orthogonal delay-Doppler division Multiplexing}
\newacronym{sbl}{SBL}{sparse Bayesian learning}
\newacronym{grasbi}{GRASBI}{grid refinement and adjustment-based sparse Bayesian inference}
\newacronym{tgraesbi}{T-GRAESBI}{grid refinement and adjustment-based sparse Bayesian inference with Student's t distribution}
\newacronym{ogsbi}{OGSBI}{off-grid sparse Bayesian inference}

\newacronym{gesbi}{GESBI}{grid evolution-based sparse Bayesian inference}
\newacronym{tgeesbi}{T-GEESBI}{grid evolution-based efficient sparse Bayesian inference with Student's t distribution}
\newacronym{1d}{1D}{one-dimensional}
\newacronym{qse}{QSE}{$q$-shift estimator}
\newacronym{mqse}{mQSE}{modified 
	\gls{qse}}
\newacronym{io}{I/O}{input-output}

\newacronym{omp}{OMP}{orthogonal matching pursuit}
\newacronym{somp}{SOMP}{structured OMP}
\newacronym{kd}{$K$D}{$K$-dimensional}
\newacronym{embb}{eMBB}{enhanced mobile broadband}
\newacronym{urllc}{URLLC}{ultra-reliable low-latency communications}
\newacronym{mmtc}{mMTC}{massive machine-type communications}

\newacronym{fembb}{feMBB}{further enhanced mobile broadband}
\newacronym{eurllc}{eURLLC}{enhanced ultra-reliable low-latency communications}
\newacronym{ummtc}{umMTC}{ultra-massive machine-type communications}
\newacronym{mimo}{MIMO}{multiple-input multiple-output}

\newacronym{ldhmc}{LDHMC}{long-distance and high-mobility communications}
\newacronym{elpc}{eLPC}{enhanced low-power communications}

\newacronym{tdma}{TDMA}{time division multiple access}
\newacronym{fdma}{FDMA}{frequency division multiple access}
\newacronym{cdma}{CDMA}{code division multiple access}

\newacronym{ber}{BER}{bit error rate}
\newacronym{awgn}{AWGN}{additive white Gaussian noise}
\newacronym{ma}{MA}{multiple access}
\newacronym{2g}{2G}{second generation}
\newacronym{3g}{3G}{third generation}
\newacronym{sms}{SMS}{short message service}
\newacronym{ofdma}{OFDMA}{orthogonal frequency division multiple access}
\newacronym{scfdma}{SC-FDMA}{single-carrier frequency division multiple access}
\newacronym{fbmc}{FBMC}{filter bank multicarrier}
\newacronym{oob}{OOB}{out-of-band}
\newacronym{fofdm}{f-OFDM}{filtered OFDM}
\newacronym{oma}{OMA}{orthogonal multiple access}
\newacronym{noma}{NOMA}{non-orthogonal multiple access}
\newacronym{qam}{QAM}{quadrature amplitude modulation}
\newacronym{dmt}{DMT}{discrete  multitone}
\newacronym{mu}{MU}{mulit-user}
\newacronym{bd}{BD}{block diagonalization}
\newacronym{zak}{ZAK}{Zak}
\newacronym{dzt}{DZT}{discrete Zak transform}
\newacronym{idzt}{IDZT}{inverse discrete Zak transform}
\newacronym{rcpotfs}{RCP-OTFS}{reduced cyclic-prefix OTFS}
\newacronym{rzpotfs}{RZP-OTFS}{reduced zero-pad OTFS}
\newacronym{cpotfs}{CP-OTFS}{cyclic-prefix OTFS}
\newacronym{dtzt}{DTZT}{discrete time Zak transform}
\newacronym{fde}{FDE}{frequency domain equalization}
\newacronym{mrc}{MRC}{maximum ratio combining}
\newacronym{map}{MAP}{maximum a posteriori}
\newacronym{zf}{ZF}{zero forcing}
\newacronym{zdf}{ZC}{Zadoff-Chu}
\newacronym{gsvd}{GSVD}{generalized singular value decomposition}
\newacronym{dfe}{DFE}{decision feedback equalizer}
\newacronym{mp}{MP}{message passing}
\newacronym{oamp}{OAMP}{orthogonal approximate message passing}
\newacronym{gmp}{GMP}{gaussian message passing}
\newacronym{mamp}{MAMP}{memory approximate message passing}
\newacronym{nmse}{NMSE}{normalized mean square error}
\newacronym{las}{LAS}{likelihood ascent search}
\newacronym{mmse}{MMSE}{minimum mean square error}
\newacronym{sp}{SP}{superimposed pilot}
\newacronym{ldpc}{LDPC}{low-density parity-check}
\newacronym{epicce}{EPICCE}{message passing (expectation propagation)-based interference canceling channel estimation}
\newacronym{ul}{UL}{uplink}
\newacronym{dl}{DL}{downlink}
\newacronym{vbl-somp}{VBL-SOMP}{variable block length-simultaneous orthogonal matching pursuit}
\newacronym{sbee}{SBEE}{smoothed basis expansion extrapolation}
\newacronym{v2i}{V2I}{vehicle-to-infrastructure}
\newacronym{rsu}{RSU}{ roadside unit}
\newacronym{ula}{ULAs}{uniform linear arrays}
\newacronym{sdr}{SDR}{software-defined radio}
\newacronym{zc}{ZC}{Zadoff–Chu}
\newacronym{sinr}{SINR}{signal-to-noise ratio}
\newacronym{egc}{EGC}{equal gain combining}
\newacronym{ccdf}{CCDF}{complementary cumulative distribution function}
\begin{document}

\title{Orthogonal Zadoff-Chu Based Pilot Design For Fractional Channel Estimation in MIMO-OTFS\\
{}
\thanks{This work was conducted within the MiFuture project, which has received funding from the European Union’s Horizon Europe (HE) Marie
	Skłodowska-Curie Actions MiFuture HORIZON-MSCA2022-DN-01, under Grant Agreement number 101119643. The work was partially supported by Portuguese national funds through FCT – Fundação para a Ciência e a Tecnologia, I.P., and, when eligible, co-funded by EU funds under project/support UID/50008/2025 – Instituto de Telecomunicações, with DOI identifier "https://doi.org/10.54499/UID/50008/2025".}
}

\author[1,2]{Omid Aghda}
\author[1,3]{Prabhat Gupta} 
\author[1,2]{João Guerreiro} 
\author[1,4]{Nuno Souto} 
\author[6]{\\Michal Szczachor} 
\author[6]{Lukasz Krzymien}
\author[1,2]{Rui Dinis}
 \affil[1]{Instituto de Telecomunicações, Lisboa, Portugal} \affil[2]{Universidade Nova de Lisboa, Monte da Caparica, 2829-516 Caparica, Portugal} 
\affil[3]{DEEC, University of Coimbra, Portugal}
\affil[4]{ ISCTE-Instituto Universitário de Lisboa,
	1649-026 Lisbon, Portugal}
\affil[6]{Nokia, Wroclaw, Poland}
%
%
\maketitle

\begin{abstract}
In this paper, we propose a Zadoff–Chu (ZC) sequence-based pilot design for fractional channel estimation in orthogonal time frequency space (OTFS) modulation for point-to-point multi-input multi-output (MIMO) scenarios. Orthogonal pilot sequences are assigned to each transmit antenna using ZF sequences. At the receiver, a correlation-based scheme is used to construct the signal for estimation. The estimation relies on a two-dimensional extension of the Aboutanios and Mulgrew (A\&M) frequency estimation method. Simulation results show that the proposed approach achieves low peak-to-average power ratio (PAPR) while maintaining stable performance as the number of transmit antennas increases.

\end{abstract}


\section{Introduction}
\Gls{6g} mobile networks introduce new use cases compared to \gls{5g} mobile communications, such as \gls{fembb}, \gls{eurllc}, \gls{ummtc}, \gls{ldhmc}, and \gls{elpc}. These use cases require further investigation in system design. In particular, \gls{ldhmc} requires reliable communication for mobile terminals operating at speeds of up to 1000~km/h, enabling scenarios such as deep-sea tourism, space tourism, and ultra-high-speed rail systems~\cite{Zegrar10700683,Aldababsa10584089,Alwis9397776}. One of the recently proposed techniques to address these challenges is \gls{otfs} modulation \cite{Hadani7925924}. \gls{otfs} aims to provide reliable communication under highly dynamic channel conditions, making it a strong candidate for meeting the stringent requirements of \gls{6g} use cases.

A major challenge in \gls{otfs} arises because signal processing for channel estimation and data detection is performed in the digital domain, where the delay and Doppler values of each wireless path may not align with integer sampling points. This misalignment, known as fractional delay and Doppler, distorts the perceived \gls{csi} and degrades the \gls{dd} channel sparsity~\cite{Guocheng11145903}, which in turn affects symbol detection performance.
Existing channel estimation methods handle this issue in different ways. Virtual-tap-based approaches assume that the observed non-integer taps correspond directly to integer-grid taps and perform estimation based on these approximations rather than the true underlying channel ~\cite{Raviteja8671740,Shen8727425,Zhao9184852}. However, as shown in \cite{AbbassiAghda2025OTFS}, this mismatch leads to degraded performance in channel equalization and data detection.
On the other hand, more advanced techniques attempt to explicitly estimate fractional delay and Doppler components. For example, the method in \cite{Wei_OGSBI_9738478} employs a dense virtual grid to first approximate the channel in the delay–Doppler domain, followed by a refinement stage based on \gls{sbl} to recover the fractional parameters by updating hyperparameters. While this approach improves estimation accuracy, it incurs high computational complexity, similar to other \gls{sbl}-based methods~\cite{Shan10857972,Shan_Gesbi_10283794}.

To address these issues, we showed in \cite{AbbassiAghda2025OTFS} that the problem of fractional delay and Doppler estimation problem in \gls{otfs} modulation can be formulated equivalently as a frequency estimation problem. Based on this insight, we applied a \gls{2d} \gls{am} method to estimate the fractional delay and Doppler parameters. However, this approach relies on impulse-based pilot transmission. A key drawback of such pilots is the high \gls{papr} of the transmitted signal, which is undesirable in practical systems~\cite{Aghda_conference_10334727,Aghda_journal_10614842}.

In this paper, we propose a precoding scheme based on \gls{zdf} sequences for a \gls{mimo}-\gls{otfs} system, where the sequences are employed as pilots in the \gls{dd} domain. By exploiting their correlation properties, the received signal can be processed using an \gls{am}-based approach to estimate the fractional delay and Doppler parameters. In this design, the pilot is spread across the entire \gls{dd} domain, which results in a lower \gls{papr} compared to the impulse-based approach in \cite{AbbassiAghda2025OTFS}.
Furthermore, by assigning \gls{zdf} sequences with different root indices to each transmit antenna, the proposed method enables simultaneous pilot transmission over the same resource grid. This allows efficient utilization of resources while maintaining separability of the signals at the receiver through correlation-based processing.

\textbf{Notation:} Boldface uppercase letters (e.g., $\mathbf{A}$) denote matrices, boldface lowercase letters (e.g., $\mathbf{a}$) denote vectors, and regular lowercase letters (e.g., $a$) denote scalars.  The $(l,k)$-th element of the matrix $\mathbf{A}$ is denoted by $\mathbf{A}[l,k]$, corresponding to the $l$-th row and $k$-th column. The indices $l$ and $k$ denote delay and Doppler indices, respectively, and take values $l = 0,\ldots,M-1$ and $k = 0,\ldots,N-1$. $\mathbf{A}(:,1)$ denotes the first column of $\mathbf{A}$. 
The Hermitian operator is denoted by $(\cdot)^{H}$,  the Kronecker product by $\otimes$, and the Hadamard product by $\odot$. The matrix $\mathbf{F}_N$ represents the $N$-point discrete Fourier transform (DFT) matrix. The notation $\sim \mathcal{CN}(\mu, \sigma^2)$ denotes a complex Gaussian random variable with mean $\mu$ and variance $\sigma^2$. Finally, $\mathbf{0}_M$ and $\mathbf{I}_M$ denote the all-zero vector of size $M$ and the $M \times M$ identity matrix, respectively.

\section{System Model}
In the \gls{otfs} transceiver, the \gls{dd} domain is divided into $M$ and $N$ bins along the delay and Doppler axes, respectively. The $l$-th delay bin ($l = 0,1,\ldots,M-1$) corresponds to a delay of $\tau = \frac{l}{M \Delta f}$, and the $k$-th Doppler bin ($k = 0,1,\ldots,N-1$) corresponds to a Doppler shift of $\nu = \frac{k}{N T_s}$, where $\Delta f$ is the subcarrier spacing. The available bandwidth is $B = M \Delta f$, and $T = N T_s$ is the frame duration, where $T_s$ is the OTFS sub-symbol duration and $T_s \Delta f = 1$.

We consider a point-to-point \gls{mimo} \gls{otfs} system with $N_t$ transmit antennas and $N_r$ receive antennas. The signal representation in the \gls{dd} domain for the $t$-th transmit antenna is denoted by $\mathbf{X}_t \in \mathbb{C}^{M \times N}$, for $t = 1,\ldots,N_t$. At the receiver, the \gls{dd}-domain signal observed at the $r$-th receive antenna is denoted by $\mathbf{Y}_r \in \mathbb{C}^{M \times N}$, for $r = 1,\ldots,N_r$.
The \gls{io} relationship between the $t$-th transmit and $r$-th receive antenna is given by
\begin{equation}
	\mathbf{y}_r = \sum_{t=1}^{N_t}\mathbf{\Phi}_t\!\left(\mathbf{l}_{\tau}, \mathbf{k}_{\nu}\right)\mathbf{h}_{r,t} + \mathbf{z}_r,
	\label{eq: sum y_r = Phi * h}
\end{equation}
where $\mathbf{h}_{r,t} = [h_{{r,t,1}}, \ldots, h_{{r,t,P}}]^T \in \mathbb{C}^{P \times 1}$ denotes the channel gain vector between the $t$-th transmit antenna and the $r$-th receive antenna, with $h_{r,t,i}, i = 1,\cdots,P$ representing the complex gain of the $i$-th propagation path, and $\mathbf{z}_r \in \mathbb{C}^{MN \times 1}\sim\mathcal{CN}(\mathbf{0}_{MN},\sigma^2\mathbf{I}_{MN})$ represents the \gls{awgn} vector at $r$-th received antenna. The vector $\mathbf{y}_r \in \mathbb{C}^{MN \times 1}$ is the received signal in the \gls{dd} domain. The matrix $\mathbf{\Phi}_t\!\left(\mathbf{l}_{\tau}, \mathbf{k}_{\nu}\right) \in \mathbb{C}^{MN \times P}$ is a function of the normalized delay and Doppler index vectors $\mathbf{l}_{\tau} = [l_{\tau_1}, \ldots, l_{\tau_P}]^T \in \mathbb{R}^{P \times 1}$ and $\mathbf{k}_{\nu} = [k_{\nu_1}, \ldots, k_{\nu_P}]^T \in \mathbb{R}^{P \times 1}$. It represents the coupling between the transmitted symbols (pilot or data) and the delay and Doppler indices associated with the $P$ propagation paths. Here, $l_{\tau_i} = \tau_i B$ and $k_{\nu_i} = \nu_i T$ denote the normalized delay and Doppler shift, respectively, where $\tau_i$ and $\nu_i$ are the physical delay and Doppler of the $i$-th path with complex gain $h_{r,t,i}$.  The normalized delay and Doppler indices can be written as $l_{\tau_i} = l_{\tau_i,I} + l_{\tau_i,F}$ and $k_{\nu_i} = k_{\nu_i,I} + k_{\nu_i,F}$, where $l_{\tau_i,I} \in \mathbb{Z} \cap [0, l_{\max}]$ and $k_{\nu_i,I} \in \mathbb{Z} \cap [-k_{\max}, k_{\max}]$ denote the integer parts, and $l_{\tau_i,F}, k_{\nu_i,F} \in [-0.5,\,0.5]$ denote the fractional parts. Here, $l_{\max} = \tau_{\max} B$ and $k_{\max} = \nu_{\max} T$ represent the maximum normalized delay and Doppler spreads of the channel, respectively. We assume that the delay and Doppler parameters are identical across all transmit–receive antenna pairs, while only the complex path gains $h_{r,t,i}$ vary with the antenna indices.
\begin{figure*}[t]
	\centering
	\includegraphics[width=0.85\linewidth]{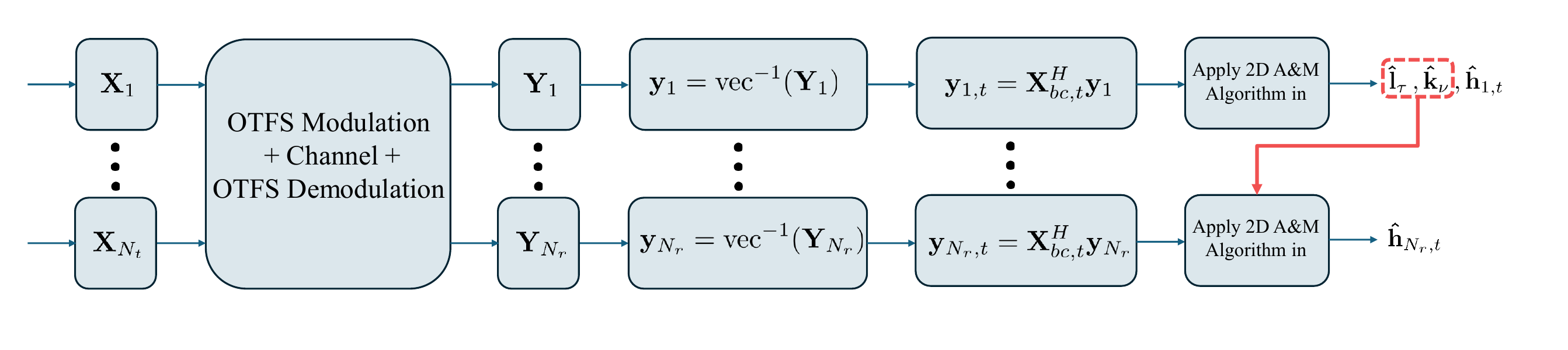}
	\caption{Block diagram of the OTFS transceiver. 
		Each transmit antenna uses an orthogonal pilot sequence with corresponding block circulant matrices $\mathbf{X}_{bc,t}$ satisfying $\mathbf{X}_{bc,t}^H \mathbf{X}_{bc,t'} = \mathbf{I}$. 
		At the receiver, the signal at the $r$-th antenna is correlated with $\mathbf{X}_{bc,t}$, followed by the A\&M algorithm for channel parameter estimation. 
		Assuming common delay and Doppler across all antenna pairs, estimates from one antenna pair are used to determine the gains of the others (shown in red).}
		\label{ref:system model}
\end{figure*}

The matrix $\mathbf{\Phi}_t\!\left(\mathbf{l}_{\tau}, \mathbf{k}_{\nu}\right)$ can be written as
\begin{equation}
	\mathbf{\Phi}_t\!\left(\mathbf{l}_{\tau}, \mathbf{k}_{\nu}\right) = \big[\boldsymbol{\phi}_t\!\left(l_{\tau_1}, k_{\nu_1}\right) \;  \cdots \; \boldsymbol{\phi}_t\!\left(l_{\tau_P}, k_{\nu_P}\right) \big],
	\label{eq: phi = [phi_1,...,phi_p]}
\end{equation}
where $\boldsymbol{\phi}_t\!\left(l_{\tau_i}, k_{\nu_i}\right) \in \mathbb{C}^{MN \times 1}$ denotes the contribution of the $i$-th propagation path characterized by normalized delay $l_{\tau_i}$ and normalized Doppler $k_{\nu_i}$.
The $(kM + l)$-th element of $\boldsymbol{\phi}_t\!\left(l_{\tau_i}, k_{\nu_i}\right)$ is given by 
\begin{equation}
	\begin{split}
		\big[\boldsymbol{\phi}_t\!\left(l_{\tau_i}, k_{\nu_i}\right)\big]_{kM + l}&= \; \sum_{k'=0}^{N-1} \sum_{l'=0}^{M-1}
	\mathbf{X}_t[k',l']\\
	&	 \, \zeta_M\!\left( l - l' - l_{\tau_i}\right)\,\zeta_N\!\left(k - k' - k_{\nu_i}\right),
	\end{split}
	\label{eq:phi = sumsum X zeta_M zeta_N}
\end{equation}
where $\zeta_{X}(x) = \frac{1-\exp(j2\pi x)}{1-\exp(j2\pi x/X)}$ is a periodic function with the period of $X$ \cite{Wei_OGSBI_9738478}.
To model the two-dimensional circular convolution in \eqref{eq:phi = sumsum X zeta_M zeta_N}, the vector $\boldsymbol{\phi}_t\!\left(l_{\tau_i}, k_{\nu_i}\right)$ can be expressed in matrix form as
\begin{equation}
	\boldsymbol{\phi}_t\!\left(l_{\tau_i}, k_{\nu_i}\right)
	= \mathbf{X}_{\text{bc,t}} \, \mathbf{w}\!\left(l_{\tau_i}, k_{\nu_i}\right),
	\label{eq:Phi = Xw}
\end{equation}
where $\mathbf{X}_{\text{bc,t}} \in \mathbb{C}^{MN \times MN}$ is a block-circulant matrix with $N$ blocks, each circulant of size $M \times M$, whose first column is given by $\mathbf{X}_{\text{bc,t}}(:,1) =\mathbf{x}_t= \mathrm{vec}(\mathbf{X}_t)$. The vector $\mathbf{w}\!\left(l_{\tau_i}, k_{\nu_i}\right) \in \mathbb{C}^{MN \times 1}$ captures the delay and Doppler shifts, thereby isolating the channel parametrization from the transmitted signal representation with entries
\begin{equation}
	\big[\mathbf{w}\!\left(l_{\tau_i}, k_{\nu_i}\right)\big]_{kM+l}
	= \zeta_M\!\left(l - l_{\tau_i}\right)\,\zeta_N\!\left(k - k_{\nu_i}\right).
\end{equation}

Replacing \eqref{eq:Phi = Xw} in \eqref{eq: phi = [phi_1,...,phi_p]} and inserting in \eqref{eq: sum y_r = Phi * h}  result in
\begin{equation}
			\mathbf{y}_r = \sum_{t=1}^{N_t}\mathbf{X}_{\text{bc,t}} \, \mathbf{W}\!\left(\mathbf{l}_{\tau}, \mathbf{k}_{\nu}\right)\mathbf{h}_{r,t} + \mathbf{z}_r 
\end{equation}
where \begin{equation}
	\mathbf{W}\!\left(\mathbf{l}_{\tau},\mathbf{k}_{\nu}\right) = \big[\mathbf{w}\!\left(l_{\tau_1}, k_{\nu_1}\right),\cdots,\mathbf{w}\!\left(l_{\tau_P}, k_{\nu_P}\right)\big]\in\mathbb{R}^{MN\times P}.
\end{equation}

\section{Pilot Design and Channel Estimation}
Before introducing the proposed pilot design, we first present the following lemmas.
\begin{lemma}
	A square block-circulant matrix $\mathbf{A}$ of size $MN \times MN$, consisting of $N$ blocks of size $M \times M$, where each block is circulant, can be diagonalized via eigen-decomposition as
	\begin{equation}
		\mathbf{A} = (\mathbf{F}_N \otimes \mathbf{F}_M)^{H} \, \mathrm{diag}(\mathbf{p}) \, (\mathbf{F}_N \otimes \mathbf{F}_M),
	\end{equation}
	where $\mathbf{p}\in\mathbb{C}^{MN\times1}$ contains the eigenvalues of $\mathbf{A}$~\cite{combescure2009block}.
\end{lemma}

\begin{lemma}
	Let $\mathbf{A} \in \mathbb{C}^{M \times M}$ be a normal matrix (i.e., $\mathbf{A}^H \mathbf{A} = \mathbf{A}\mathbf{A}^H$). 
	If all eigenvalues $\lambda$ of $\mathbf{A}$ satisfy $|\lambda| = 1$, then
	$
	\mathbf{A}^H \mathbf{A} = \mathbf{I}
	$ \cite{strang2016linear}.
\end{lemma}
\subsection{Pilot Design}
	Based on Lemma~1 and Lemma~2, we use the  \gls{zdf} sequence, which is widely adopted in 5G due to its favorable correlation properties, as the eigenvalue vector of the transmitted pilot symbol to construct the block-circulant matrix $\mathbf{X}_{\text{bc},t}$, i.e.,
	\begin{equation}
		\mathbf{X}_{\text{bc},t} =\sqrt{MN} (\mathbf{F}_N \otimes \mathbf{F}_M)^{H} \, \mathrm{diag}(\mathbf{p}_t) \, (\mathbf{F}_N \otimes \mathbf{F}_M),
		\label{eq:Xbc = (FN Kron FM)^H diag(pt)(FN Kron FM)}
	\end{equation}
	where $\mathbf{p}_t \in \mathbb{C}^{MN \times 1}$ denotes the \gls{zc} sequence associated with the $t$-th transmit antenna~\cite{3gpp_ts36211_v16}. With this construction, the resulting matrix $\mathbf{X}_{\text{bc},t}$ is orthogonal, i.e, $\mathbf{X}_{\text{bc},t}^H\mathbf{X}_{\text{bc},t}=\mathbf{I}$.
	In \eqref{eq:Xbc = (FN Kron FM)^H diag(pt)(FN Kron FM)}, the scaling factor $\sqrt{MN}$ is introduced to ensure that the transmitted signal has unit power, i.e., $\mathbb{E}\left\lbrace|\mathbf{X}_t[l,k]|^2\right\rbrace=1$.
	
	Considering that $N_t < MN$, which is a practical assumption, distinct ZC sequences with different root indices can be assigned to the transmit antennas \cite{3gpp_ts36211_v16}. 
	
	\subsection{Pilot Correlation}
		To estimate the channel between the $t$-th transmit antenna and the $r$-th receive antenna, we exploit the orthogonality of the matrix $\mathbf{X}_{bc,t}$. By projecting the received signal onto this orthogonal basis, we obtain
		\begin{equation}
			\begin{split}
				\mathbf{X}_{bc,t}^H \mathbf{y}_r =& \mathbf{X}_{bc,t}^H\mathbf{X}_{bc,t} \mathbf{W}\!\left(\mathbf{l}_{\tau}, \mathbf{k}_{\nu}\right)\mathbf{h}_{r,t} \\
				&+ \sum_{\substack{t^\prime=1,\\t^\prime\not=t}}^{N_t} \mathbf{X}_{\text{bc,t}}^H \mathbf{X}_{bc,t^\prime}\, \mathbf{W}\!\left(\mathbf{l}_{\tau}, \mathbf{k}_{\nu}\right)\mathbf{h}_{r,t^\prime} + \mathbf{X}_{bc,t}^H\mathbf{z}_r \\
				=&MN\mathbf{W}\!\left(\mathbf{l}_{\tau}, \mathbf{k}_{\nu}\right)\mathbf{h}_{r,t}\\
				&+ \sum_{\substack{t^\prime=1,\\t^\prime\not=t}}^{N_t}  \mathbf{X}_{bc,t,t^\prime}\, \mathbf{W}\!\left(\mathbf{l}_{\tau}, \mathbf{k}_{\nu}\right)\mathbf{h}_{r,t^\prime} + \mathbf{z}_{r,t}
			\end{split}
			\label{eq: Xy = IWh + XXWh + z}
		\end{equation}
		where, based on Lemma~1, $\mathbf{X}_{bc,t,t^\prime} \in \mathbb{C}^{MN \times MN}$ is a block-circulant matrix whose eigenvalues are given by $\mathbf{p}_{t,t^\prime} = \mathbf{p}_t \odot \mathbf{p}_{t^\prime} \in \mathbb{C}^{MN \times 1}$, and $\mathbf{z}_{r,t} = \mathbf{X}_{bc,t}^H \mathbf{z}_r \in \mathbb{C}^{MN \times 1}\sim\mathcal{CN}(\mathbf{0}_{MN},MN\sigma^2\mathbf{I}_{MN})$ denotes the transformed noise vector.
		The second term in \eqref{eq: Xy = IWh + XXWh + z} corresponds to inter-antenna interference affecting the estimation of the channel between the $t$-th transmit antenna and the $r$-th receive antenna. This term is absorbed into the noise term and  \eqref{eq: Xy = IWh + XXWh + z}  becomes
		\begin{equation}
			\mathbf{y}_{r,t}=\mathbf{X}_{bc,t}^H \mathbf{y}_r = MN\mathbf{W}\!\left(\mathbf{l}_{\tau}, \mathbf{k}_{\nu}\right)\mathbf{h}_{r,t} + \mathbf{\bar{z}}_{r,t}\in\mathbb{C}^{MN\times 1}
			\label{eq:Xy = MNWh + z}
		\end{equation}
			where $\mathbf{\bar{z}}_{r,t}$ denotes the aggregate effect of inter-antenna interference and noise.

	\subsection{Channel Estimation}
		In \eqref{eq:Xy = MNWh + z}, by neglecting noise and interference, each element of $\mathbf{y}_{r,t}$ is given by 
		\begin{equation}
			\big[\mathbf{y}_{r,t}\big]_{kM+l} = MN\sum_{i=1}^{P}h_{{r,t,i}}\zeta_M\!\left(l - l_{\tau_i}\right)\,\zeta_N\!\left(k - k_{\nu_i}\right).
			\label{eq:yrt = sum MN h_i zetaM zetaN}
		\end{equation}
		In \cite{AbbassiAghda2025OTFS},  it is shown that when the received samples take the form of \eqref{eq:yrt = sum MN h_i zetaM zetaN}, the problem of estimating the channel delay and Doppler admits the same solution structure as a frequency estimation problem. Consequently, conventional low-complexity frequency estimation methods, such as the \gls{am} algorithm, can be applied. In this work, we employ the \gls{2d} version of the \gls{am} algorithm to estimate the delay and Doppler parameters, and subsequently project the received signal onto the estimated delay and Doppler basis to estimate the channel gains. 	
		
		Assuming the number of taps \(P\) is known\footnote{Following \cite{Serbes_multiple_frequency8977344}, we assume the number of channel taps is known, which simplifies the estimation process. When this information is unavailable, the method proposed in \cite{Li485924, rissanen1983universal, Diao6670774, Roy32276, Schmidt1143830} can be used to estimate the number of taps effectively.
			In practical scenarios where the number of taps is not specified, established model order selection techniques such as the generalized Akaike information criterion (GAIC) or minimum description length (MDL) can be applied as a preprocessing step to estimate the number of taps, as discussed in \cite{Li485924, rissanen1983universal}.}, the algorithm runs for \(P\) external iterations, with each iteration estimating the parameters \(\hat{l}_\tau^{(p)}\), \(\hat{k}_\nu^{(p)}\), and \(h^{(p)}_{{r,t}}\) of one channel tap.\footnote{Note that the index $i$ in \eqref{eq:yrt = sum MN h_i zetaM zetaN} refers to the true channel taps, while superscript $p$ here denotes the iteration index of the estimation algorithm, which does not necessarily correspond to the ordering of the true taps.}
		At $p$-th iteration
		first, we reshape the signal $\mathbf{y}_{r,t}^{(p)}$ into an $(M,N)$ matrix $\mathbf{Y}_{r,t}^{(p)} \in \mathbb{C}^{M \times N}$, with initialization  $\mathbf{y}_{r,t}^{(0)} = \mathbf{y}_{r,t}$.  We then identify the delay and Doppler indices corresponding to the peak magnitude of the received signal as
		\begin{equation}
			\hat{l}_{\tau_{I}}^{(p)}, \hat{k}^{(p)}_{\nu_{I}} = \operatorname*{argmax}_{l,k} |\mathbf{Y}^{(p-1)}_{r,t}[l,k]|
			\label{eq:find max integer delay and doppler}
		\end{equation}
		where $\hat{l}_{\tau_{I}}^{(p)}$ and $\hat{k}_{\nu_{I}}^{(p)} $ denote the estimated integer delay and Doppler at $p$-th iteration. The reliability of this peak detection is valid only when the received \gls{snr} is above the \gls{ml} threshold \cite{Aboutanios1408177,Serbes8977344}. 
		Here, we describe the fractional Doppler estimation procedure; the fractional delay estimation follows the same approach, with the corresponding dual variables and parameters associated with delay and Doppler appropriately exchanged \cite{AbbassiAghda2025OTFS}.

		To estimate the fractional part, the algorithm runs for \(J\) internal iterations, during which the fractional Doppler \( \hat{k}_{\nu,F}^{(p,j)} \) is obtained by interpolating the row corresponding to the estimated integer delay \( \hat{l}_{\tau_I}^{(p)} \) to evaluate the signal at \( \hat{k}_{\nu,I}^{(p)} \pm \alpha \). At, $j$-th iteration, the interpolation results in
		\begin{equation}
			\begin{split}
						y^{(p,j)}_{r,t,\pm\alpha,\nu} =& \frac{1}{N}\sum_{k=0}^{N-1}\mathbf{Y}^{(p-1)}_{r,t}[\hat{l}_{\tau,{I}}^{(p)},k]\\
						& \zeta_M\!\left( k-\hat{k}_{\nu,{I}}^{(p)}-\hat{k}_{\nu,{F}}^{(p,j-1)} \pm\alpha\right)
			\end{split}
			\label{eq:yrtalpha}
		\end{equation}
		where $\hat{k}_{\nu,{F}}^{(p,0)}=0$, and $\alpha=0.5$.
		Then, the fractional Doppler is updated as
		\begin{equation}
			\hat{k}_{\nu,{F}}^{(p,j)} = \alpha\, \Re\left\{ \frac{y^{(p,j)}_{r,t,+\alpha,\nu} + y^{(p,j)}_{r,t,-\alpha,\nu}}{y^{(p,j)}_{r,t,+\alpha,\nu} - y^{(p,j)}_{r,t,-\alpha,\nu}} \right\} + \hat{k}_{\nu,{F}}^{(p,j-1)} 
			\label{eq:delay estimation fractional estimation}
		\end{equation}
		Finally, the Doppler estimate at the \(p\)-th iteration is given by $\hat{k}_{\nu}^{(p)} = \hat{k}_{\nu,{I}}^{(p)} + \hat{k}_{\nu,{F}}^{(p,J)}$. Using the same procedure with dual parameters and values we estimate the delay at $p$-th iteration as $\hat{l}_{\tau}^{(p)} = \hat{l}_{\tau,{I}}^{(p)} + \hat{l}_{\tau,{F}}^{(p,J)}$.
		
		Now, to estimate the channel gain $h_{r,t}^{(p)}$ we project the $\mathbf{Y}_{r,t}^{(p-1)}$ into the estimated delay and Doppler $\hat{l}_{\tau}^{(p)}$ and $\hat{k}_{\nu}^{(p)}$ as
		\begin{equation}
			\begin{split}
				h_{r,t}^{(p)} =& \frac{1}{(MN)^2}\sum_{l=0}^{M-1}\sum_{k=0}^{N-1}\mathbf{Y}_{r,t}^{(p-1)}[l,k]\\
				&  \zeta_M\!\left( l-\hat{l}_{\tau}^{(p)}\right)\,\zeta_N\!\left(k-\hat{k}_{\tau}^{(p)}\right) 
			\end{split}
			\label{eq:hrthat}
		\end{equation}
		And at the end we update $\mathbf{y}_{r,t}^{(p-1)}$ to cancel  the contribution of $p$-th estimated path as
		\begin{equation}
			\mathbf{y}_{r,t}^{(p)} = \mathbf{y}_{r,t}^{(p-1)} - MN	h_{r,t}^{(p)}\mathbf{w}\!\left(\hat{l}_{\tau}^{(p)},\hat{k}_{\nu}^{(p)}\right)
			\label{eq:cancel contribution}
		\end{equation}
		
		The algorithm is summarized in Algorithm~1, and the block diagram of the proposed transceiver is shown in Fig.~\ref{ref:system model}.
		Since the delay and Doppler shift of each tap are assumed to be common across all antenna pairs, using only one antenna-pair can save complexity, look at the red part in Fig.~\ref{ref:system model}, but performance could be improved if all the antenna pairs contributed to the estimation. 
		This can be extended  to the general case where the delays of the taps can vary between antenna pairs. The obtained estimates are then used for gain estimation across all antenna pairs. Joint estimation using combining methods such as \gls{mrc} or \gls{egc} may further improve accuracy.
		
\begin{algorithm}[t]
	\caption{Channel Estimation Algorithm for the Proposed Pilot Design}\label{algo1}
	\begin{algorithmic}[1]
		\Require $\mathbf{y}_{r,t};\; \hat{k}_{\nu,F}^{(0)} = 0;\; \hat{l}_{\tau,F}^{(0)} = 0$.
		\Ensure $\hat{{l}}^{(p)}_{\tau};\hat{{k}}^{(p)}_{\nu};\hat{{h}}^{(p)}_{r,t}$ for $p=1,\cdots,P$.
		\For{$p=1,...,P$}
		\State  $ 	\hat{l}_{\tau_{I}}^{(p)}, \hat{k}^{(p)}_{\nu_{I}}\Leftarrow  $  Equation \eqref{eq:find max integer delay and doppler}
		\For{$j=1,...,J$}
		\State $y^{(p,j)}_{r,t,\pm\alpha,\nu},y^{(p,j)}_{r,t,\pm\alpha,\tau} \Leftarrow$ Equation \eqref{eq:yrtalpha} and its dual.
		\State $ \hat{k}_{\nu,{F}}^{(p,j)}, \hat{l}_{\tau,{F}}^{(p,j)}\Leftarrow$ Equation \eqref{eq:delay estimation fractional estimation} and its dual.
		\EndFor
		\State $\hat{k}_{\nu}^{(p)} = \hat{k}_{\nu,{I}}^{(p)} + \hat{k}_{\nu,{F}}^{(p,j)}$, $\hat{l}_{\tau}^{(p)} =\hat{l}_{\tau,{I}}^{(p)} + \hat{l}_{\tau,{F}}^{(p,j)}$
		\State  $ 	h_{r,t}^{(p)}\Leftarrow  $  Equation \eqref{eq:hrthat}
		\State  $ 	\mathbf{y}_{r,t}^{(p)}\Leftarrow  $  Equation \eqref{eq:cancel contribution}
		\EndFor 
	
	\end{algorithmic}
\end{algorithm}
		
%
%
		
\section{Simulation Results}
The performance of the proposed method is evaluated via MATLAB simulations. 
The carrier frequency is $4\,\text{GHz}$, and a $P=4$ path channel is considered with $l_{\max}=4$ and $k_{\max}=3$, where delays and Dopplers are uniformly distributed. 
A total of 3000 Monte Carlo iterations are performed with $M=29,N=23$, and the subcarrier spacing is $\Delta f = 15\,\text{kHz}$. 
The pilot power is set such that $\mathrm{SNR} = \frac{\mathbb{E}[|\mathbf{X}[l,k]|^2]}{\sigma^2}$. Number of internal iteration for the A\&M algorithm is set to $J=2$ \cite{Aboutanios1408177}.
\begin{figure}[!t]
	\centering
		\includegraphics[width=0.9\columnwidth]{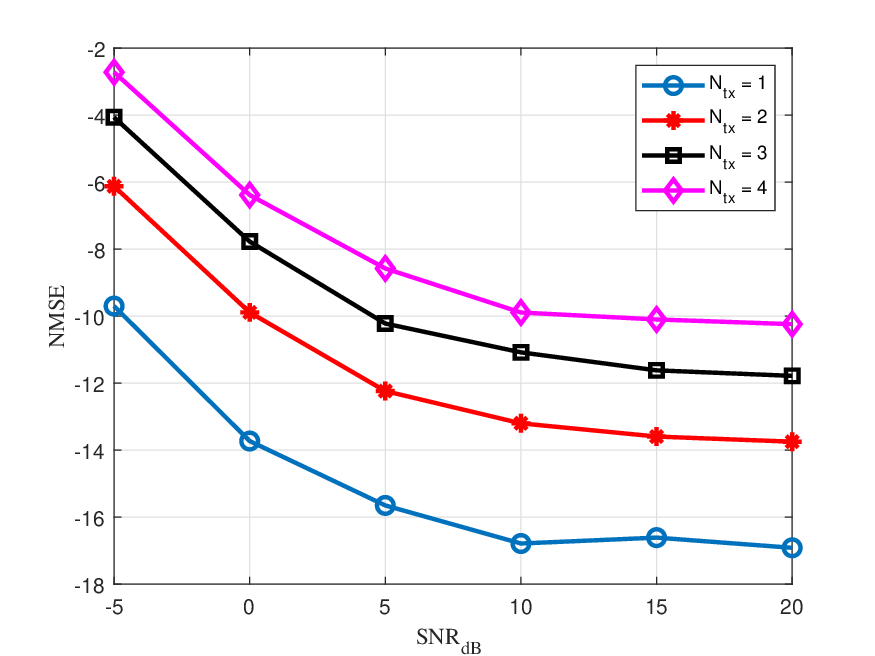}
		\caption{ NMSE versus SNR.}
		\label{fig:nmse}
\end{figure}
\begin{figure}[!t]
	\centering
		\includegraphics[width=0.9\columnwidth]{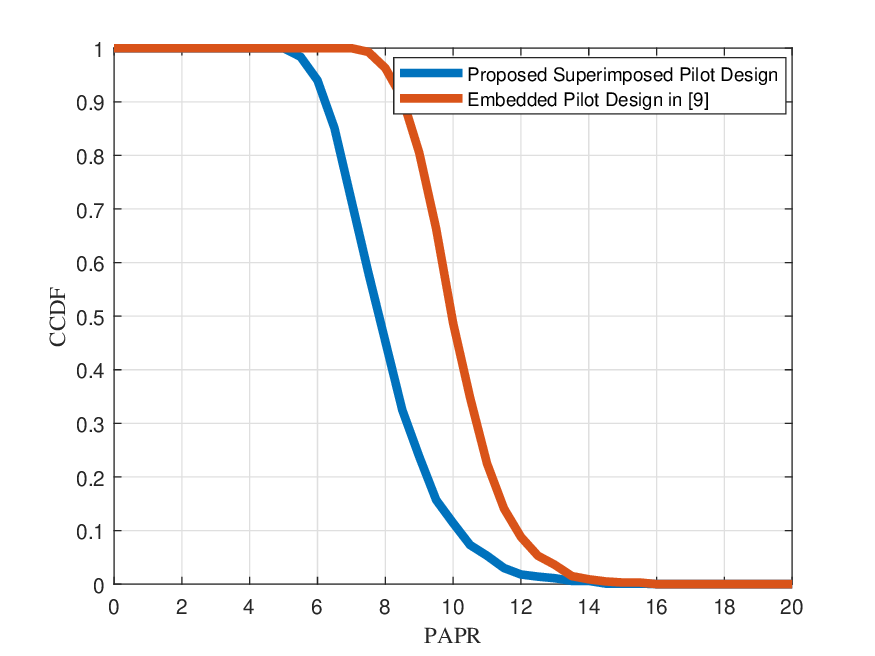}
		\caption{CCDF of PAPR}
		\label{fig:papr}
\end{figure}
In Fig.~\ref{fig:nmse}, the \gls{nmse} is plotted versus the \gls{snr} for different numbers of antennas. The estimation criterion is defined as
\begin{equation}
	\text{NMSE} = 
	\frac{\left\|\mathbf{W}\!\left(\mathbf{l}_{\tau}, \mathbf{k}_{\nu}\right)\mathbf{h}_{r,t} 
		- \mathbf{W}\!\left(\hat{\mathbf{l}}_{\tau}, \hat{\mathbf{k}}_{\nu}\right)\hat{\mathbf{h}}_{r,t}\right\|_F^2}
	{\left\|\mathbf{W}\!\left(\mathbf{l}_{\tau}, \mathbf{k}_{\nu}\right)\mathbf{h}_{r,t}\right\|_F^2}.
\end{equation}
As expected, increasing the number of transmit antennas leads to higher interference power, which degrades the channel estimation performance.

In Fig.~\ref{fig:papr}, the \gls{ccdf} of the \gls{papr} for the transmitted signal is shown. As discussed, the proposed method achieves a lower \gls{papr} compared to the pilot design and channel estimation scheme in~\cite{AbbassiAghda2025OTFS}. For a fair comparison, data is included in both methods. In our approach,  data is superimposed on the pilot such that, at each resource element $\mathbf{X}[l,k]$, the transmitted signal is the sum of the data and pilot components.	 To maintain the total transmitted signal power equal to 1, the pilot power ratio is set to 0.17. In~\cite{AbbassiAghda2025OTFS}, the pilot power is chosen as $(l_{\max}+1)(2k_{\max}+1)$ to allocate dedicated zero-valued guard regions as pilot support, which results in the pilot occupying approximately 0.17 of the total transmit power.

\section{conclusion}
In this paper, we proposed a \gls{zdf}-based pilot design for channel estimation in \gls{mimo}-\gls{otfs} systems. A mathematical formulation of the pilot design for each transmit antenna was provided. The results demonstrate that the proposed method achieves a lower \gls{papr} compared to the embedded pilot-based channel estimation scheme.

In future work, we are looking to propose an integration of the proposed pilot design with data, such that the pilot can coexist with data transmission. In addition, we plan to exploit the combination of received antenna signals to improve the accuracy of delay and Doppler estimation.

\bibliography{vtc2026references}

@ARTICLE{Raviteja8671740,
	author={Raviteja, P. and Phan, Khoa T. and Hong, Yi},
	journal={IEEE Transactions on Vehicular Technology}, 
	title={Embedded Pilot-Aided Channel Estimation for {OTFS} in Delay–Doppler Channels}, 
	year={2019},
	volume={68},
	number={5},
	pages={4906-4917},
	doi={10.1109/TVT.2019.2906357}}

@ARTICLE{Aboutanios1408177,
	author={Aboutanios, E. and Mulgrew, B.},
	journal={IEEE Transactions on Signal Processing}, 
	title={Iterative frequency estimation by interpolation on Fourier coefficients}, 
	year={2005},
	volume={53},
	number={4},
	pages={1237-1242},
	doi={10.1109/TSP.2005.843719}}

@ARTICLE{Serbes_multiple_frequency8977344,
	author={Serbes, Ahmet and Qaraqe, Khalid},
	journal={IEEE Signal Processing Letters}, 
	title={A Fast Method for Estimating Frequencies of Multiple Sinusoidals}, 
	year={2020},
	volume={27},
	number={},
	pages={386-390},
	doi={10.1109/LSP.2020.2970837}}

@ARTICLE{Zegrar10700683,
	author={Zegrar, Salah Eddine and Sümer, Ahmet Sacid and Arslan, Hüseyin},
	journal={IEEE Transactions on Vehicular Technology}, 
	title={Fractional Delay and Fractional Doppler Estimation and Mitigation Framework in OTFS Systems}, 
	year={2025},
	volume={74},
	number={2},
	pages={2884-2896},
	doi={10.1109/TVT.2024.3470839}}

@ARTICLE{Alwis9397776,
	author={Alwis, Chamitha De and Kalla, Anshuman and Pham, Quoc-Viet and Kumar, Pardeep and Dev, Kapal and Hwang, Won-Joo and Liyanage, Madhusanka},
	journal={IEEE Open Journal of the Communications Society}, 
	title={Survey on {6G} Frontiers: Trends, Applications, Requirements, Technologies and Future Research}, 
	year={2021},
	volume={2},
	number={},
	pages={836-886},
	doi={10.1109/OJCOMS.2021.3071496}}

@ARTICLE{Aldababsa10584089,
	author={Aldababsa, Mahmoud and Özyurt, Serdar and Kurt, Güneş Karabulut and Kucur, Oğuz},
	journal={IEEE Open Journal of the Communications Society}, 
	title={A Survey on Orthogonal Time Frequency Space Modulation}, 
	year={2024},
	volume={5},
	number={},
	pages={4483-4518},
	doi={10.1109/OJCOMS.2024.3422801}}

@INPROCEEDINGS{Hadani7925924,
	author={Hadani, R. and Rakib, S. and Tsatsanis, M. and Monk, A. and Goldsmith, A. J. and Molisch, A. F. and Calderbank, R.},
	booktitle={2017 IEEE Wireless Communications and Networking Conference (WCNC)}, 
	title={Orthogonal Time Frequency Space Modulation}, 
	year={2017},
	volume={},
	number={},
	pages={1-6},
	doi={10.1109/WCNC.2017.7925924}}

@ARTICLE{Serbes8977344,
	author={Serbes, Ahmet and Qaraqe, Khalid},
	journal={IEEE Signal Processing Letters}, 
	title={A Fast Method for Estimating Frequencies of Multiple Sinusoidals}, 
	year={2020},
	volume={27},
	number={},
	pages={386-390},
	doi={10.1109/LSP.2020.2970837}}

@ARTICLE{Shan10857972,
	author={Shan, Yaru and Shafie, Akram and Yuan, Jinhong and Wang, Fanggang},
	journal={IEEE Transactions on Wireless Communications}, 
	title={Off-Grid Channel Estimation for Orthogonal Delay-Doppler Division Multiplexing Using Grid Refinement and Adjustment}, 
	year={2025},
	volume={24},
	number={4},
	pages={3590-3605},
	doi={10.1109/TWC.2025.3532459}}

@ARTICLE{Shen8727425,
	author={Shen, Wenqian and Dai, Linglong and An, Jianping and Fan, Pingzhi and Heath, Robert W.},
	journal={IEEE Transactions on Signal Processing}, 
	title={Channel Estimation for Orthogonal Time Frequency Space ({OTFS}) Massive MIMO}, 
	year={2019},
	volume={67},
	number={16},
	pages={4204-4217},
	doi={10.1109/TSP.2019.2919411}}

@ARTICLE{Aghda_journal_10614842,
	author={Aghda, Omid Abbassi and Omidi, Mohammad Javad and Saeedi-Sourck, Hamid},
	journal={IEEE Transactions on Vehicular Technology}, 
	title={Low-PAPR Joint Channel Estimation and Data Detection in {ZP-OTFS} System}, 
	year={2024},
	volume={73},
	number={12},
	pages={19821-19825},
	doi={10.1109/TVT.2024.3435934}}

@INPROCEEDINGS{Aghda_conference_10334727,
	author={Aghda, Omid Abbassi and Omidi, Mohammad Javad and Saeedi-Sourck, Hamid},
	booktitle={2023 31st International Conference on Electrical Engineering (ICEE)}, 
	title={Superimposed Channel Estimation in {OTFS} Modulation Using Compressive Sensing}, 
	year={2023},
	volume={},
	number={},
	pages={756-761},
	doi={10.1109/ICEE59167.2023.10334727}}

@ARTICLE{Zhao9184852,
	author={Zhao, Lei and Gao, Wen-Jing and Guo, Wenbin},
	journal={IEEE Communications Letters}, 
	title={Sparse Bayesian Learning of Delay-Doppler Channel for {OTFS} System}, 
	year={2020},
	volume={24},
	number={12},
	pages={2766-2769},
	doi={10.1109/LCOMM.2020.3021120}}

@ARTICLE{Wei_OGSBI_9738478,
	author={Wei, Zhiqiang and Yuan, Weijie and Li, Shuangyang and Yuan, Jinhong and Ng, Derrick Wing Kwan},
	journal={IEEE Transactions on Wireless Communications}, 
	title={Off-Grid Channel Estimation With Sparse Bayesian Learning for {OTFS} Systems}, 
	year={2022},
	volume={21},
	number={9},
	pages={7407-7426},
	doi={10.1109/TWC.2022.3158616}}

@INPROCEEDINGS{Shan_Gesbi_10283794,
	author={Shan, Yaru and Wang, Fanggang and Hao, Yaxing},
	booktitle={2023 IEEE International Conference on Communications Workshops (ICC Workshops)}, 
	title={Off-Grid Channel Estimation Using Grid Evolution in Rectangular Pulse-Shaping {OTFS} System}, 
	year={2023},
	volume={},
	number={},
	pages={295-300},
	doi={10.1109/ICCWorkshops57953.2023.10283794}}

@ARTICLE{Li485924,
	author={Li, Jian and Stoica, P.},
	journal={IEEE Transactions on Signal Processing}, 
	title={Efficient mixed-spectrum estimation with applications to target feature extraction}, 
	year={1996},
	volume={44},
	number={2},
	pages={281-295},
	doi={10.1109/78.485924}}

@article{rissanen1983universal,
	title={A universal prior for integers and estimation by minimum description length},
	author={Rissanen, Jorma},
	journal={The Annals of statistics},
	volume={11},
	number={2},
	pages={416--431},
	year={1983},
	publisher={Institute of Mathematical Statistics}
}

@ARTICLE{Schmidt1143830,
	author={Schmidt, R.},
	journal={IEEE Transactions on Antennas and Propagation}, 
	title={Multiple emitter location and signal parameter estimation}, 
	year={1986},
	volume={34},
	number={3},
	pages={276-280},
	doi={10.1109/TAP.1986.1143830}}

@ARTICLE{Roy32276,
	author={Roy, R. and Kailath, T.},
	journal={IEEE Transactions on Acoustics, Speech, and Signal Processing}, 
	title={ESPRIT-estimation of signal parameters via rotational invariance techniques}, 
	year={1989},
	volume={37},
	number={7},
	pages={984-995},
	doi={10.1109/29.32276}}

@ARTICLE{Diao6670774,
	author={Diao, Ruipeng and Meng, Qingfeng},
	journal={IEEE Transactions on Instrumentation and Measurement}, 
	title={An Interpolation Algorithm for Discrete Fourier Transforms of Weighted Damped Sinusoidal Signals}, 
	year={2014},
	volume={63},
	number={6},
	pages={1505-1513},
	doi={10.1109/TIM.2013.2289585}}

@article{combescure2009block,
	title={Block-circulant matrices with circulant blocks, Weil sums, and mutually unbiased bases. II. The prime power case},
	author={Combescure, Monique},
	journal={Journal of mathematical physics},
	volume={50},
	number={3},
	year={2009},
	publisher={AIP Publishing}
}

@book{strang2016linear,
	author    = {Gilbert Strang},
	title     = {Introduction to Linear Algebra},
	edition   = {5},
	publisher = {Wellesley-Cambridge Press},
	year      = {2016},
	isbn      = {9780980232776}
}

@misc{3gpp_ts36211_v16,
	author       = {{3GPP}},
	title        = {{Evolved Universal Terrestrial Radio Access (E-UTRA); Physical channels and modulation}},
	howpublished = {{3GPP TS 36.211 version 16.2.0 Release 16}},
	year         = {2020},
	note         = {Available online: https://www.3gpp.org}
}

@article{AbbassiAghda2025OTFS,
	author  = {Omid Abbassi Aghda and Oussama Ben Haj Belkacem and João Guerreiro and et al.},
	title   = {{A\&M} Based Fractional Delay and Doppler Channel Estimation in {OTFS}},
	journal = {Research Square},
	year    = {2025},
	note    = {Preprint, Version 1, August 31, 2025},
	doi     = {10.21203/rs.3.rs-7381395/v1},
	url     = {https://doi.org/10.21203/rs.3.rs-7381395/v1}
}

@ARTICLE{Guocheng11145903,
	author={Guo, Cheng and Jiao, Jian and Bai, Siyuan and Wu, Ziyang and Wang, Ye and Zhang, Qinyu},
	journal={IEEE Internet of Things Journal}, 
	title={{OTFS}-Based Super Resolution Channel Estimation in Multi-Satellite Coordinated Transmission}, 
	year={2025},
	volume={},
	number={},
	pages={1-1},
	doi={10.1109/JIOT.2025.3604565}}
\bibliographystyle{IEEEtran}



\end{document}